\documentclass[twocolumn,showpacs,preprintnumbers,amsmath,amssymb]{revtex4}
\usepackage{graphicx}
\input epsf
\usepackage{dcolumn}
\usepackage{bm}

\begin{document}


\title{The nonsingular brane solutions via the Darboux transformation}

\author{A.V. Yurov}
\email{artyom_yurov@mail.ru}
\author{V.A. Yurov}%
 \email{yurov@freemail.ru}
\affiliation{%
The Theoretical Physics Department, Kaliningrad State
University,A. Nevskogo str., 14, 236041,
 Russia.
\\
}%


\date{\today}

\begin{abstract}
We consider the Darboux transformation as a method of construction
of exact nonsingular solutions describing the three-dimensional
brane that interacts with five-dimensional gravity and the bulk
scalar field. To make it work, the five-dimensional Einstein's
equations and the Israel's  conditions are being reduced to the
Schr\"odinger equation with the jump-like potential and the wave
functions sewing conditions in jump point correspondingly. We show
further that it is always possible to choose the functions in
Crum's determinants in such way, that the five-dimensional Ricci
scalar $R$ will always be finite both on brane and in bulk. The
new exact solutions being the generalizations of the model with
the odd superpotential are presented. Described formalism is also
appliable to the cases of more realistic branes with cosmological
expansion. As an example, via the usage of the simple orbifold
model ($S_1/{\Bbb Z}_2$) and one-time Darboux transformation we
construct the models where the cosmological constant on the
visible brane is exponentially small.
\end{abstract}

\pacs{11.25.Mj, 11.27.+d}
\maketitle

\section{\label{sec:level1}Introduction
}

In article \cite{Flanagan}, there have been considered an
interesting model, which describes the set of parallel 3-branes,
all of them being imbedded in five-dimensional bulk, filled with
the gravitation and the scalar field. The action of set is
expressed as:
\begin{equation}
\begin{array}{c}
S=\int d^4xdy\sqrt{\mid
g\mid}\left(\frac{1}{2}R-\frac{(\nabla\phi^2)}{2}-V(\phi)\right)-\\
\sum_b\int_{y_b} d^4x\sqrt{\mid{\tilde g}^b\mid}\sigma_b(\phi),
\label{action}
\end{array}
\end{equation}
 where $(x^{\mu},y)$ and $0\le\mu\le 3$ are the
five-dimensional coordinates; brane number $b$ is located at
$y=y_b$; $g_{ab}$ is the 5-D metric, and ${\tilde g}^b_{\mu\nu}$
is the metric, induced on the $b$th brane. (Note that it is
written in special system of units, where the gravitational
constant $\kappa=1$ and speed of light $c=1$.) The tension on the
$b$th brane is named $\sigma_b$, and the potential $V(\phi)$ is
considered to be a function of the bulk scalar field
$\phi=\phi(y)$. The field equations follows from the
(\ref{action}) by the procedure of variation; their solution is
assumed to have a form (\cite{Flanagan}):
\begin{equation}
ds^2=dy^2+A(y)\left(-dt^2+e^{2Ht}\delta_{ik}dx^idx^k\right),
\label{interval}
\end{equation}
where $A(y)$ is a sought warp factor, and $H$=const is a Hubble
parameter. If $H=0$ {\em on the brane}, the brane will be the
stationary one, and if $H>0$, then the brane will be expanding in
the de Sitter regime \cite{uskor}.

The (\ref{action}) leads to system of differential equations,
that, in turn, can be rewritten in "supersymmetric" style by
introduction of the superpotential $W(\phi)$, being defined by the
${\dot\phi}\equiv W'(\phi)/2$ relation, where the dot denotes the
derivative in $y$, and the prime -- the derivative in $\phi$. In
case of $H=0$ there exists following equation
\begin{equation}
V(\phi)=W'^2/8-W^2/6. \label{superpotential}
\end{equation}

Upon usage of this equation, the authors of \cite{Flanagan} have
suggested a simple way of obtaining of exact solutions of the
field equations (see also \cite{RS}, \cite{TW} for further
details). Their approach can be expressed in following steps:
\newline
(i) Choosing convenient $V(\phi)$ and $W(\phi)$, making sure that
the equation (\ref{superpotential}) with these new values still
holds true;
\newline
(ii) Integration of equation ${\dot\phi}\equiv W'(\phi)/2$, in
order to receive the relation $\phi=\phi(y)$;
\newline
(iii) When $H=0$ there exists the relation ${\dot
A}(y)/A(y)=-W(\phi(y))/3$. Hence, substitution of the obtained
function for $\phi(y)$ in explicit equation for $W(\phi)$ and
further integration results in the sought warp factor $A(y)$.

From this point, the famous solutions RS (Randall-Sundrum) turns
out to be just a case $W(\phi)=\pm a^2={\rm const}$. Authors
\cite{Flanagan} have also considered the models with even
($W(\phi)=\pm\left(a^2-b\phi^2\right)$), odd ($W(\phi)=\pm
2b^2\phi$) and exponential ($W(\phi)=2a{\rm e}^{-k\phi}$)
potentials (the even superpotentials had been previously discussed
in \cite{DFGK}, and the odd ones had been introduced in
\cite{kaku}).

The major advantage of the described method lies in it's
simplicity. For example, it allows an easy construction of
nonsingular solutions (see also \cite{kaku}). Lets consider the
case of one brane, localized in $y=y_0$. From the equation
${\dot\phi}\equiv W'(\phi)/2$ it is possible to express $y$ as a
function of the field variable $\phi$:
\begin{equation}
y(\phi)=y_0+\int_{\phi_0}^{\phi}\frac{d{\tilde \phi}}{W'({\tilde
\phi})}, \label{y(phi)}
\end{equation}
where $\phi_0=\phi(y_0)$ is a value of a scalar field on brane. It
follows from the (\ref{y(phi)}) that the very existence of zeros
of the function $W'(\phi)$ ($\phi=\phi^{\pm}={\rm const}$,
$\mid\phi^{\pm}\mid <\infty$) is leading to the divergences of the
integral. In another words, $\phi\to\phi^{\pm}$ when
$y\to\pm\infty$. What it means, is that if we choose the
superpotential as having at least two zeros at both $\phi>\phi_0$
and $\phi<\phi_0$ where $W'$ is vanishing fast enough (for example
$W'(\phi)\sim\mid\phi-\phi^{\pm}\mid^k$, with $k\ge 1$, and
$\phi\to\phi^{\pm}$), then the field $\phi$ remains finite
throughout all the $bulk$ space. On the other hand, the case of
$H=0$ allows to express the Ricci scalar via the superpotential
and its derivative in the form:
\begin{equation}
R(\phi)=2(W'(\phi))^2/3-5W^2(\phi)/9, \label{R}
\end{equation}
thus, showing the finiteness of the scalar curvature for the
aforementioned nonsingular superpotential for every $y$. The
particular examples of the nonsingular solutions for even, odd and
exponential superpotentials can all be found in \cite{Flanagan}.
It also contains the generalization of this technique for the
non-stationary branes ($H\ne 0$) and a brief discussion touching
the possible usages of the method in solving such a well known
problems as the problem of cosmological constant's smallness and
the hierarchy problem.

However, in spite of all the strong points, this method does have
its apparent weaknesses, the primal one being its
"phenomenological" nature - the fact, that each physically
consistent (i.e. leading to the reasonable $\phi(y)$ and $A(y)$)
superpotential $W(\phi)$ can only be obtained by "guessing".
Although all simple examples, considered in \cite{Flanagan} were
not too hard "to guess" and they do lead to sensible cosmological
models, it would not necessarily be that easy, whenever we try the
more difficult (and hence, more realistic) $W(\phi)$. These
reasonings lead us to principal question: does there exist such
procedure of (\ref{action})'s nonsingular solutions construction,
which is both mathematically simple and regular? In this work we
give the positive answer for this question and introduce the new
method, which satisfies both of the conditions.

The general idea of our approach can be formulated as follows: the
supersymmetric link between $V(\phi)$ and $W(\phi)$ is a typical
one for the supersymmetric quantum mechanics (SSQM). In turn, SSQM
is realizable through the usage of Darboux transformation (DT) for
the Schr\"odinger equation \cite{Salle}, \cite{ABI}, and that
allows us to construct the integrable nonsingular potentials -
providing that all the prop functions of Crum's determinants do
have the intermittent zeros \cite{Salle}, \cite{Adler}. So, it
seems to be only natural trying to write the DT for the field
equations directly, and then using them as a source of a new exact
solutions. An existence of quite simple algebraic connections
between the Ricci scalar and the superpotential (\ref{R}) notes
that construction of nonsingular potentials in SSQM via the right
DT application will also aid in development of the systematical
procedure of construction of the nonsingular solutions
($R\ne\infty$) in the brane theory, which, by that means, will
turn out to be a regular one - as distinct from the
"phenomenological approach" of \cite{Flanagan}. In particular, the
DT method, if applied to just a single exactly solvable potential,
grants us the infinite number of an exact nonsingular solutions,
with the mentioned potential being a "fuse" in all those solutions
construction. For example, in this work we'll take the RS for the
role of such a fuse solution. Another interesting way of the DT
usage is a construction of the chains of discrete symmetries and
the further inquiries of their closures. Such a technique allows
getting the new exact solutions, which are manifested through the
Painlev\'e transcendents and the higher extensions.

The article has the following structure. Next paragraph is
dedicated to the Darboux transformations and to their
compatibility with the Israels jump conditions at the brane
\cite{Israel}. In third paragraph we'll present some new exact
solutions, received by one-time Darboux dressing  of the RS
solutions. Double DT, the importance of the shape-invariance and
the dressing chains are all examined in the fourth paragraph. In
particular, it will be shown, that even, odd and exponential
superpotentials, discussed in work \cite{Flanagan} are just the
examples of the shape-invariants with regard to DT potentials. In
the same section we'll show the exact solutions, that serve as the
generalizations of the model with the odd superpotential (here we
include the solutions expressed via the fourth Painlev\'e
transcendent). In paragraph 5 we'll describe the construction of
an exact solutions for the non-stationary branes, made with the
usage of DT and will discuss the nascent difficulties. Here we'll
consider both noncompact and compact spacetimes. In the last case,
using the simple orbifold model ($S_1/{\Bbb Z}_2$) and one-time
Darboux transformation we'll construct the models with two branes
where the cosmological constant on the one of them (visible brane)
will be exponentially small. Also, we'll outline the new approach
(called Green function method) which allows to construct solutions
both in stationary and non-stationary cases. Finally, the six
paragraph is an overall conclusion.

\section{Darboux transformations and jump conditions}

If $H=0$ and $u\equiv {\dot A}/A$, then from the (\ref{action})
and (\ref{interval}) we get the following set of equations
\cite{Flanagan}: \begin{equation} {\dot
u}=-\frac{2{\dot\phi}^2}{3}-\frac{2}{3}\sum_b\sigma_b(\phi)\delta(y-y_b),\qquad
u^2=\frac{{\dot\phi}^2}{3}-\frac{2}{3}V(\phi). \label{dvaurav}
\end{equation}
Third equation has the form
$$
\ddot\phi+2u\dot\phi=\frac{\partial
V(\phi)}{\partial\phi}+\sum_b\frac{\partial\sigma_b(\phi)}{\partial\phi}\delta(y-y_b),
$$
and is not independent, because it can be received from the set
(\ref{dvaurav}). In case of the single brane, located at $y=0$, we
get $\sigma(\phi(y))\delta(y)=\sigma_0\delta(y)$, where
$\sigma_0=\sigma(\phi(0))$ (supposing $\phi(0)\ne 0$). It is easy
to check, that the function $A^2(y)$ satisfies the Schr\"odinger
equation
\begin{equation}
\left(\frac{d^2}{dy^2}+\frac{8}{3}V(\phi(y))+\beta\delta(y)\right)A^2(y)=0,
\label{shr}
\end{equation}
where $\beta=4\sigma_0/3$.
Now lets define the spectral problem
\begin{equation}
\ddot\psi=\left(v(y)-\beta\delta(y)-\lambda\right)\psi,
\label{schrodinger}
\end{equation}
where $\psi=\psi(y;\lambda)$ and $v(y)$ is a certain potential
which will be further refereed to as a quantum-mechanical
potential, or QM-potential. Lets assume, that it is possible to
solve the equation (\ref{schrodinger}) formally (i.e., obtaining
among others some "non-physical" solutions which do not belong to
$L^2$) for some QM-potential $v(y)$ and all $\lambda\in \Bbb R$.
Lets then suppose that for some values of the spectral parameter
$\lambda$ (say, $\lambda=\lambda_0$) the solutions of the
(\ref{schrodinger}) will be strictly positive:
$\psi(y;\lambda_0)>0$. In this case we can conclude that function
$A(y;\lambda_0)=\sqrt{\psi(y;\lambda_0)}$ is the solution of the
(\ref{shr}) with potential
\begin{equation}
V(\phi(y))=\frac{3}{8}\left(\lambda_0-v(y)\right).
\label{V}
\end{equation}
Therefore, the equation (\ref{schrodinger}) can, in fact, be
considered as the producer of exact solutions of the (\ref{shr}).
Substituting $A(y)$ and $V(\phi(y))$, obtained from the
(\ref{schrodinger}), in (\ref {dvaurav}) we'll find out
$\phi=\phi(y)$. In this point there can be raised the question:
what solutions can be named the physically sensible one? More
precisely, what can be said about the solutions with
${\dot{\phi}}^2<0$?
\newline
\newline
{\bf Nota Bene 1}. Recently the great deal of interest has been
drawn to those very specifical cosmological models, which admits
the violation of the dominant-energy condition \cite{1}. Such
models often presuppose that the "dark energy" has so-called
phantom component \cite{2}, which should be described by some
scalar field with the negative kinetic term. Quantum theory of
those fields has been considered in \cite{3} (this work also has
an almost complete bibliography of a subject). As a general result
of this investigations we can point out the following indication:
apparently such fields fails to be consistently considered as a
fundamental objects. But nevertheless, we can't totally exclude
the probability that the Lagrangians with the negative kinetic
terms will somehow appear in a role of effective models. For
example, similar terms appear in some models of supergravity
\cite{4} and in the gravity theories with highest derivatives
\cite{5}. Finally, the "phantom energy" in the brane theory was
considered in \cite{7}. However, despite of all the above said, we
will not further consider the phantom component. In other words,
we shall suppose that ${\dot\phi}^2>0$ during the rest of this
paper.
\newline

For specificity, we'll further assume the following condition for
the potential $v(y)$:
\begin{equation}
\displaystyle{ v(y)={\rm const}+{\tilde v}(y),\qquad
\int_{-\infty}^{+\infty}dy\,\left(1+|y|\right)|{\tilde
v}(y)<\infty.} \label{potential}
\end{equation}
It is also worth mentioning, that for our further investigation
the validity of (\ref{potential}) is not really necessary. We just
found it to be convenient to consider the potentials, that are
confined within such bounds.

So, lets assume the parameters $\psi(y)$, $v(y)$, $\lambda$ to be
already determined. We are going to find the functions $A(y)$,
$\phi(y)$ and $V(\phi)$. In order to do this we'll examine the
region located far from the brane to exterminate the
$\delta$-function term in the (\ref{schrodinger}) from further
considerations. It's obvious, that the quantity $A(y)$ can only be
restored if $\psi(y)$ is nonnegative for all $y$. The only one
wave function that both satisfies this condition and belongs to
the physical spectrum of Schr\"odinger equation
(\ref{schrodinger}), is a wave function of ground state with
$\lambda=\lambda_{_{vac}}$.
We should also mention, that there exists another possible way to
choose the "good" wave function, namely, to consider the case
$\lambda=\lambda_{-1}<\lambda_{_{vac}}$. This satisfies the
imposed condition, since the (\ref{schrodinger}) indeed admits the
always positive solutions $\psi(y;\lambda_{-1})$. These solutions
obviously do not belong to the $L_2$ space and definitely do not
describe the bound states, but the warp factor
$A=\sqrt{\psi(y;\lambda_{_1})}$ will nevertheless be real.
However, these solutions result in singularity at infinity:
$A\to\infty$ at $y\to\infty$, and therefore the ground state
$\psi(y;\lambda_{_{vac}})$ is by far the best alternative to
$\psi(y;\lambda_{-1})$

It is well known, that DT has a remarkable capacity to enable the
engineering of ad hoc potentials with arbitrary finite discrete
spectrum, thus it can be consistently considered as a procedure of
construction of the system (\ref{dvaurav})'s exact solutions.
We'll now quickly remind the very essence of DT method
\cite{Darboux}. Let $\psi_1=\psi(y;\lambda_1)$ and
$\psi_2=\psi(y;\lambda_2)$ be the two solutions of the equation
(\ref{schrodinger}), and let also $\psi_1>0$ for every $y$ it is
defined on. The $\psi_1$ is called the prop function of Darboux
transformation. In general, the transformation law is:
\begin{equation}
\begin{array}{cc}
\displaystyle{\psi_2\to
\psi_2^{(1)}=\frac{{\dot\psi_2}\psi_1-{\dot\psi_1}\psi_2}{\psi_1}},\\
\\
\displaystyle{v\to v^{(1)}=v-2\frac{d^2}{dy^2}\log\psi_1, \qquad
\lambda_2\to\lambda_2}. \label{dt}
\end{array}
\end{equation}
Darboux transformation (\ref{dt}) is an isospectral symmetry of
equation (\ref{schrodinger}), because the dressed function
$\psi_2^{(1)}$ is the solution of the dressed equation
(\ref{schrodinger}):
$$
\ddot\psi_2^{(1)}=\left(v^{(1)}-\beta^{(1)}\delta(y)-\lambda_2\right)\psi_2^{(1)},
$$
with a new (dressed) potential $v^{(1)}$ but with the same value
of $\lambda^{(1)}_2=\lambda_2$. The transformation law for the
prop function is given by the relation (\cite{Salle}):
\begin{equation}
\begin{array}{c}
\psi_1\to\psi_1^{(1)}=\frac{1}{\psi_1}\left(C_1+C_2\int
\psi_1^2(y) dy\right),\\
\\
\ddot\psi_1^{(1)}=\left(v^{(1)}-\beta^{(1)}\delta(y)-\lambda_1\right)\psi_1^{(1)},
\label{selfdt}
\end{array}
\end{equation}
with arbitrary constants $C_{1,2}$ (and, of course,
$\lambda^{(1)}_1=\lambda_1$). The most interesting part for us
here is a ground state of a dressed Hamiltonian. There exist three
possibilities.
\newline

1) If $\psi_1$ is a wave function of a ground state, and
$\lambda_2$ is an energy of a first excited level, then the new
dressed function $\psi^{(1)}_2$ will be the wave function of a new
Hamiltonian's ground state \cite{ABI}. All other values of the
discrete spectrum of this Hamiltonian are received by simple
obliteration of level $\lambda_1$ from the initial spectrum.
\newline

2) The prop function $\psi_1=\psi_1(y;\lambda_1)$ does not belong
to an $L_2$, but it doesn't has any zeros either, and $\psi_1\to
e^{+p^2 |y|}$ when $y\to\pm \infty$. In the case, after the DT
we'll get the new Hamiltonian, whose discrete spectrum is just an
old one except for the new state $\lambda=\lambda_1$ added,
thereby being the new ground state. The corresponding wave
function $\psi^{(1)}_1$ (\ref{selfdt}) with $C_2=0$ becomes the
ground state's wave function \cite{bereza}.
\newline

3) The aforementioned prop function can in fact be represented as
a linear superposition of two positive, linearly independent
solutions $\psi_1^{(\pm)}$ with the following characteristics:
$\psi_1^{(\pm)}\to 0$ when $y\to\mp\infty$ and $\psi_1^{(\pm)}\to
e^{+p^2 |y|}$ when $y\to\pm \infty$. Taking either $\psi_1^{(+)}$
or $\psi_1^{(-)}$ as a prop function results in creation of a new
Hamiltonian, whose discrete spectrum is totally coincides with the
initial one \cite{bereza}.

Our article is dedicated in general to the second case.

As a next step of our consideration we should examine the jump
conditions, existing on the brane. These conditions can be written
as
\begin{equation}
\psi_{i+}(0)=\psi_{i-}(0)=\psi_i(0),\qquad {\dot\psi}_{i+}(0)-
{\dot\psi}_{i-}(0)=-\beta\psi_i(0). \label{jump}
\end{equation}
where $\psi_{i+}(y)$ and $\psi_{i-}(y)$ are the solutions of
Schr\"odinger equation (\ref{schrodinger}), both with the same
potential and eigenvalues, but valid only for positive or negative
$y$ correspondingly:
\begin{equation}
{\ddot\psi}_{i\pm}=\left(v(y)-\beta\delta(y)-\lambda_i\right)\psi_{i\pm}.
\label{schrodinger2}
\end{equation}
It is easily verified, that (\ref{jump}) are exactly the same jump
conditions as in before-cited article \cite{Flanagan}:
$$
\begin{array}{c}
 u_+(0)-u_-(0)=\frac{1}{2}\left(\frac{{\dot\psi}_+(0)}{\psi_+(0)}-\frac{{\dot\psi}_-(0)}{\psi_-(0)}\right)=\\
\frac{{\dot\psi}_+(0)-{\dot\psi}_-(0)}{2\psi(0)}=-\frac{\beta}{2}=-\frac{2\sigma_0}{3},
\end{array}
$$
being, in essence, the Israel conditions \cite{Israel},
\cite{Rubakov}. The next theorem proves, that DT keeps the jump
conditions (\ref{jump}) unchanged:
\newline
\newline
{\bf Theorem}. If
$\psi_i(y)=\left\{\psi_{i+}(y>0),\psi_{i-}(y<0)\right\}$ with
$i=1,2$ are solutions of (\ref{schrodinger2}), satisfying the
(\ref{jump}) and if
$\psi_1=\left\{\psi_{_{1+}},\psi_{_{1-}}\right\}$ is a prop
function, then the transformed function
$\psi^{(1)}_2=\left\{\psi^{(1)}_{2+},\psi^{(1)}_{2-}\right\}$
(where the prop functions of $\psi_{_{2\pm}}$ are $\psi_{_{1\pm}}$
correspondingly) satisfies the same conditions (\ref{jump}) with
$\beta^{(1)}=-\beta$:
\begin{equation}
\begin{array}{cc}
\displaystyle{\psi^{(1)}_{2+}(0)=\psi^{(1)}_{2-}(0)\equiv\psi^{(1)}_2(0)},\\
\\
\displaystyle{{\dot\psi}^{(1)}_{2+}(0)-{\dot\psi}^{(1)}_{2-}(0)=+\beta\psi_i(0)}.
\label{jump2}
\end{array}
\end{equation}

The proving of the theorem is conducted by the direct calculation.
We'll just cite two formulas, which turned to be quite useful in
proving:
$$
{\dot\tau}_{1\pm}=v-\lambda_1-\tau_{1\pm}^2,\qquad
{\dot\psi}^{(1)}_{2\pm}=\left(\lambda_1-\lambda_2\right)\psi_{_{2\pm}}-\tau_{1\pm}\psi^{(1)}_{2\pm},
$$
where  $\tau_{1\pm}={\dot\psi}_{1\pm}/\psi_{_{1\pm}}$. An obvious
result of this theorem is the conclusion, that n-times repeated DT
leads to the dressed functions, satisfying the (\ref{jump}) with
tension $\beta^{(n)}=(-1)^n\beta$, on the assumption that $n$ prop
functions-solutions $\{\psi_{i\pm}\}$ ($i=1,..,n$) of the initial
Schr\"odinger equation (\ref{schrodinger2}) and one transformed
solution $\psi_{n+1\pm}=\psi$ with the same potential, but
different $\lambda$ are all satisfy the condition (\ref{jump}).
Such n-times dressing is usually convenient to introduce via the
Crum's formulas \cite{Crum}:
\begin{equation}
A^{(n)}_{\pm}(y)=\sqrt{\frac{D^{(\pm)}_n(y)}{\Delta^{(\pm)}_n(y)}},\qquad
v^{(n)}_{\pm}(y)=v-2\frac{d^2}{dy^2}\log\Delta^{(\pm)}_n(y),
\label{crum}
\end{equation}
where
$$
\Delta^{(\pm)}_n(y)=\left| \begin{array}{cccc}
\psi^{_[n-1]}_{_{n\pm}}&\psi^{_[n-2]}_{_{n\pm}} &...&\psi_{_{n\pm}}\\
\psi^{_[n-1]}_{_{n-1\pm}}&\psi^{_[n-2]}_{_{n-1\pm}} &...&\psi_{_{n-1\pm}}\\
\star\\
\star\\
\psi^{_[n-1]}_{_{2\pm}}&\psi^{_[n-2]}_{_{2\pm}} &...&\psi_{_{2\pm}}\\
\psi^{_[n-1]}_{_{1\pm}}&\psi^{_[n-2]}_{_{1\pm}}
&...&\psi_{_{1\pm}} \end{array} \right|,
$$
$$
D^{(\pm)}_n(y)= \left|\begin{array}{ccccc}
\psi^{_[n]}_{_{\pm}}&\psi^{_[n-1]}_{_{\pm}}&\psi^{_[n-2]}_{_{\pm}}&...&\psi_{_{\pm}}\\
\psi^{_[n]}_{_{n\pm}}&\psi^{_[n-1]}_{_{n\pm}}&\psi^{_[n-2]}_{_{n\pm}} &...&\psi_{_{n\pm}}\\
\psi^{_[n]}_{_{n-1\pm}}&\psi^{_[n-1]}_{_{n-1\pm}}&\psi^{_[n-2]}_{_{n-1\pm}} &...&\psi_{_{n-1\pm}}\\
\star\\
\star\\
\psi^{_[n]}_{_{2\pm}}&\psi^{_[n-1]}_{_{2\pm}}&\psi^{_[n-2]}_{_{2\pm}} &...&\psi_{_{2\pm}}\\
\psi^{_[n]}_{_{1\pm}}&\psi^{_[n-1]}_{_{1\pm}}&\psi^{_[n-2]}_{_{1\pm}}&...&\psi_{_{1\pm}}
\end{array}
\right|,
$$
and $\psi^{[k]}\equiv d^k\psi/dy^k$. These formulas are highly
useful for studying. In particular, the solution, written in this
form, taken along with (\ref{jump2}), allows us to extend the
conditions, introduced by \cite{Flanagan}:
$$
u^{(n)}_+(0)-u^{(n)}_-(0)=\left(-1\right)^{n+1}\frac{2\sigma_0}{3},
$$
where $u^{(n)}_{\pm}={\dot A}^{(n)}_{\pm}/A^{(n)}_{\pm}$ and
$\sigma_0$ is a tension on the initial brane. Moreover, using the
(\ref{crum}) in the procedure of levels addition one can show,
that the wave function of ground state for potential
$v^{(n)}_{\pm}$ is defined by the formula:
\begin{equation} \psi^{(n)}_{_{n\pm}}=\left(A^{(n)}_{\pm}\right)^2=\frac{\Delta^{(\pm)}_{n-1}}{\Delta^{(\pm)}_n}. \label{vacuum}
\end{equation}
Wave function of a ground state is nowhere turns to zero, and this
statement is similar to condition $A^{(n)}_{\pm}\ne 0$. Using the
$n$-times dressed scale factor, we can calculate the corresponding
Ricci scalar:
\begin{equation}
R^{(n)}_{\pm}=-\frac{4 {\ddot A}^{(n)}_{\pm} A^{(n)}_{\pm}+
\left({\dot A}^{(n)}_{\pm}\right)^2}{( A^{(n)}_{\pm})^2},
\label{Rn}
\end{equation}
where the $\pm$ signs are referring to $y>0$ (over the brane) and
$y<0$ (under the brane). Since the denominator of this expression
is always non-zero, it follows right from the (\ref{vacuum}) and
(\ref{Rn}) that Ricci scalar remains finite not just on brane, but
also in a bulk. If, for example, the behavior of
$\psi^{(n)}_{_{n\pm}}$ is an asymptotical one, i.e. if
$\psi^{(n)}\to {\rm e}^{-p^2\mid y\mid}$ when $y\to \pm\infty$,
then $R^{(n)}_{\pm}\to -5p^4$ at infinity. Thus, in the case of
single stationary brane, the DT method allows the simple
construction of an abundant set of an exact nonsingular solutions
of equation (\ref{dvaurav}).
\newline
\newline
{\bf Nota Bene 2}. As we have seen, the equation
(\ref{schrodinger}) can be extremely effective in producing of the
exact solutions of (\ref{dvaurav}). In the cases of exactly
solvable potentials $v(y)$, the usage of DT allows the
construction of a large (in fact, infinite) set of such solutions.
The problem arises if the considering Schr\"odinger equation
starts out from some arbitrary solution of the (\ref{dvaurav}).
This solution gives us, in turn, the single exact solution of the
(\ref{shr}) with $v(y)=-8V(\phi(y))/3$ and $\lambda=0$ -- but,
generally, it is not guaranteed, that $v(y)=-8V(\phi(y))/3$ will
be one of exactly solvable potentials, and hence -- not provided
that we can solve the spectral problem (\ref{schrodinger}) for all
other admitted $\lambda$. This means, that, in order to use the DT
method, we should in first place carefully choose the initial
solution of the (\ref{dvaurav}). It is highly remarkable, that the
famous RS model actually does the trick!
\section{One-time dressing of the RS brane.}
To illustrate the effectiveness of formulas (\ref{dt}),
(\ref{selfdt}) we'll take the example of the RS brane dressing,
which case corresponds to a simple quantum potential
$v=\mu^2=\;$const. Here ${\tilde v}=0$ (see (\ref{potential})),
and the potential of a scalar field $V=-3\mu^2/8$. Solution
$\psi_1$ of equation (\ref{schrodinger}) with eigenvalue
$\lambda_1=0$ has the form
\begin{equation}
\begin{array}{c}
\psi_{_{1\pm}}=a_{_{1\pm}} e^{\mu y}+b_{_{1\pm}} e^{-\mu y},\qquad
\psi_{_{1+}}=\psi_1(y>0), \\ \psi_{_{1-}}=\psi_1(y<0),
\label{opor}
\end{array}
\end{equation}
where  $a_{_{1\pm}}$, and $b_{_{1\pm}}$ are real positive
constants. Usage of (\ref{jump}) results in following equations:
\begin{equation}
a_{_{1-}}=a_{_{1+}}+\frac{\beta(a_{_{1+}}+b_{_{1+}})}{2\mu},\qquad
b_{_{1-}}=b_{_{1+}}-\frac{\beta(a_{_{1+}}+b_{_{1+}})}{2\mu}.
\label{link}
\end{equation}
It is obvious from (\ref{link}) that
$a_{_{1+}}+b_{_{1+}}=a_{_{1-}}+b_{_{1-}}$, therefore it is
convenient to take advantage of the parameterization:
\begin{equation}
a_{_{1\pm}}=r^2\sin^2\alpha_{_{\pm}},\qquad
b_{_{1\pm}}=r^2\cos^2\alpha_{_{\pm}}. \label{represent}
\end{equation}
(\ref{represent}) and (\ref{link}) allows us to get the
compatibility conditions:
\begin{equation}
-\sin^2\alpha_{_{+}}<\frac{\beta}{2\mu}<\cos^2\alpha_{_{+}},
\label{compactibility}
\end{equation}
where $|\beta|<2\mu$ necessarily. Thus, in order to get the new
exact solutions via DT, starting out from the RS solutions, we
should use (\ref{opor}) combined with addition conditions
(\ref{link}) and (\ref{compactibility}).

Now, let's take (\ref{opor}) as a prop function. After the DT's
(\ref{dt}) execution we get
$$
\mu^2-\beta\delta(y)\to \mu^2+v^{(1)}_{\pm}-\beta^{(1)}\delta(y),
$$
where
$$
\displaystyle{ v^{(1)}_{\pm}=-\frac{8\mu^2
a_{_{1\pm}}b_{_{1\pm}}}{\left(a_{_{1\pm}} e^{\mu y}+ b_{_{1\pm}}
e^{-\mu y}\right)^2},}
$$
while $v^{(1)}_+= v^{(1)}(y>0)$ and $v^{(1)}_-= v^{(1)}(y<0)$.
This potential suffers a jump on the brane:
$$
v^{(1)}_+(0)-v^{(1)}_-(0)=-4\mu\beta\left(
\frac{\beta}{2\mu}+\frac{a_{_{1+}}-b_{_{1+}}}{a_{_{1+}}+b_{_{1+}}}\right),
$$
which is equally zero if the tension is
\begin{equation}
\beta=\frac{2\mu(b_{_{1+}}-a_{_{1+}})}{a_{_{1+}}+b_{_{1+}}}.
\label{nastr}
\end{equation}
This case is really special for it allows to get the potential
$v_-^{(1)}$ from $v_+^{(1)}$ by simple permutation of $a_{_{1+}}$
and $b_{_{1+}}$. Let's choose
$\psi^{(1)}_{_{1\pm}}=1/\psi_{_{1\pm}}$ \cite{thereby}. It is
clear, that for $y\to 0$:
$$
{\dot\psi}^{(1)}_{_{1+}}(0)-
{\dot\psi}^{(1)}_{_{1-}}(0)=-\beta^{(1)}\psi^{(1)}_1(0)=+\beta\psi^{(1)}_1(0),
$$
which, according to (\ref{jump2}) is exactly the way it had to be.
Using the $\psi^{(1)}_{_{1\pm}}$ we receive a new metric
$$
A^{(1)}_{\pm}=\frac{1}{\sqrt{a_{_{1\pm}} e^{\mu y}+b_{_{1\pm}}
e^{-\mu y}}}.
$$
It's asymptotical behavior is: $A^{(1)}(y)\to e^{-\mu y/2}/{\sqrt
{a_{_{1+}}}},$ when $y\to+\infty,$ and $A^{(1)}(y)\to e^{\mu
y/2}/{\sqrt{b_{_{1-}}}},$ when $y\to-\infty$. (\ref{nastr}) is a
particular case of $a_{_{1+}}=b_{_{1-}}$, thus
$$
A^{(1)}(y)\to \frac{1}{\sqrt{a_{_{1+}}}}e^{-\mu\mid y\mid/2},
\qquad y\to \pm\infty.
$$
If we additionally insert condition $b_{_{1+}}=a_{_{1-}}=0$, we
will get, after the substitution $\beta\to-\beta$, the reduction
of $A^{(1)}$ to the solution of the RS set (\ref{dvaurav}), with
the potential being equivalent to the initial $V=-3\mu^2/8$.

Returning to the general case, we receive
$$
\displaystyle{ \left({\dot\phi}^{(1)}_{\pm}\right)^2=\frac{3\mu^2
a_{_{1\pm}}b_{_{1\pm}}}{\left(a_{_{1\pm}} e^{\mu y}+ b_{_{1\pm}}
e^{-\mu y}\right)^2},}
$$
and, after the simple calculation, we come to well-known
sine-Gordon potential:
$$
V^{(1)}(\phi_{\pm})=-\frac{3\mu^2}{8}\cos\left[\frac{4}{\sqrt{3}}\left(\phi_{\pm}(y)-\phi_{0\pm}\right)\right],
$$
where
$$
\phi(y)_{\pm}=\phi_{0\pm}+\sqrt{3}\arctan\left(\sqrt{\frac{a_{_{1\pm}}}{b_{_{1\pm}}}}e^{\mu
y}\right),
$$
and $\phi_{0\pm}$ are the arbitrary constants.

Finally, to make the picture complete, we'll derive the quantity
$\left(\sigma_0^{(1)}\right)'$:
$$
\begin{array}{cc}
\left(\sigma_0^{(1)}\right)'={\dot\phi}_+^{(1)}(0)-{\dot\phi}_-^{(1)}(0)=\\
\mu\sqrt{3}\sin\left(\alpha_{_+}-\alpha_-\right)\cos\left(\alpha_{_+}+\alpha_-\right).
\end{array}
$$
Note, that in the case (\ref{nastr}) we have
$$
\beta=2\mu\cos 2\alpha_+,\qquad
\alpha_-=\alpha_++\frac{\pi}{2}\left(2N+1\right),
$$
where $N$ is a whole number.

\section{Further development of the method}
This paragraph is dedicated to three questions, each of them being
connected with the further development and generalization of the
method; namely: the $n=2$ dressing of the RS solutions, the role
of the shape-invariance and the two interesting ways of
generalization of the model with the odd superpotential. The first
way is realizable via the Adler theorem whereas for the second one
the dressing chains of discrete symmetries should be used.

\subsection{\label{sec:level2}Double dressing of the RS model}

Let's return to the RS model. The solution of it's Schr\"odinger
equation with eigenvalue $\lambda_2=\mu^2-\nu^2$ can be chosen in
form:
\begin{equation}
\begin{array}{c}
\psi_{_{2\pm}}=a_{_{2\pm}} e^{\nu y}+b_{_{2\pm}} e^{-\nu y},\qquad
\psi_{_{2+}}=\psi_1(y>0), \\ \psi_{_{2-}}=\psi_1(y<0).
\label{dressing}
\end{array}
\end{equation}

Let's assume
\begin{equation}
a_{_{2-}}=a_{_{2+}}+\frac{\beta(a_{_{2+}}+b_{_{2+}})}{2\nu},\qquad
b_{_{2-}}=b_{_{2+}}-\frac{\beta(a_{_{2+}}+b_{_{2+}})}{2\nu}.
\label{link2}
\end{equation}
Using (\ref{dt}) for (\ref{dressing}) we get
\begin{widetext}
\begin{equation}
\begin{array}{cc}
\displaystyle{
\psi^{(1)}_{_{2\pm}}=\frac{\Omega_{\pm}}{a_{_{1\pm}} e^{\mu y}+b_{_{1\pm}} e^{-\mu y}},}\\
\\ \Omega_{\pm}=(\nu-\mu)\left(a_{_{1\pm}}a_{_{2\pm}}e^{(\mu+\nu)y}-b_{_{1\pm}}b_{_{2\pm}}e^{-(\mu+\nu)y}\right)+ (\nu+\mu)\left(a_{_{2\pm}}b_{_{1\pm}}e^{(\nu-\mu)y}-a_{_{1\pm}}b_{_{2\pm}}e^{(\mu-\nu)y}\right).
\end{array}
\label{dressed}
\end{equation}
\end{widetext}
Function (\ref{dressed}) will be used as a prop function for
second DT. For the construction of a new ground state with energy
$\lambda_2<0$ we should choose $\nu>\mu>0$, $a_{_{2\pm}}>0$ and
$b_{_{2\pm}}<0$ -- this way the function $\psi^{(1)}_{_{2\pm}}$
will not have the unwanted zeros. Consideration of (\ref{link2})
gives us the conclusion, that value of positive coefficient
$a_{_{2\pm}}$ should satisfy the inequalities:
$$
\frac{a_{_{2+}}}{a_{_{2-}}}>\left(\frac{2\nu}{2\nu+\beta}\right)_{{\rm
при}\,\,\beta>0},\qquad
\frac{a_{_{2-}}}{a_{_{2+}}}>\left(\frac{2\nu}{2\nu-\beta}\right)_{{\rm
при}\,\,\beta<0}.
$$
If these relations are correct, we are free to use (\ref{crum})
and (\ref{vacuum}) for the second DT, which results in:
$$
v^{(2)}_{\pm}=\mu^2-2\frac{d^2}{dy^2}\log\Omega_{\pm},\qquad
A^{(2)}_{\pm}=\sqrt{\frac{a_{_{1\pm}} e^{\mu y}+b_{_{1\pm}}
e^{-\mu y}}{\Omega_{\pm}}}.
$$
It is quite easy to make sure, that what we get is a brane with
correct jump condition and the tension $\sigma^{(2)}_0=+\sigma_0$.
Also note, that potential $v^{(1)}$ has two levels: $\lambda_1=0$
and $\lambda_2=\mu^2-\nu^2$.

\subsection{\label{sec:level3}Shape-invariant potentials}
It appears, that the proper work of DT method is only provided for
exactly solvable potentials. One of the efficient ways of
obtaining of such potentials lies in a usage of so-called shape
invariants \cite{Infeld}. If an initial potential is a function of
$y$ and some free parameters $a_i$ : $v=v(y;a_i)$, and after the
DT one gets $v^{(1)}=v^{(1)}(y;a^{(1)}_i)$ then $v$ is called the
shape-invariant (SP) potential. SP-potentials are quite common in
quantum mechanics, e.g. the harmonic oscillator \cite{Salle}. The
major point here is that exactly solvable potentials from
\cite{Flanagan} (for the models without cosmological expansion)
are also the SP-potentials. We'll confine ourselves with
considering the three examples from the cited article, retaining
the terminology.
\newline
{\em A. Exponential potential.} Superpotential is $W(\phi)=2b
e^{-\phi}$, where $\phi(y)=\log(a-by)$. Therefore
$$
v(y)=\frac{c}{(by-a)^2},
$$
which is a well known SP potential: $v^{(1)}={\rm const}\times v$.
\newline
{\em  B. Even superpotential.} For the DFGK model \cite{DFGK} we
get (with $A(0)=1$):
$$
\log A(y)=-\frac{ay}{3}+g\left(1-e^{-2by}\right).
$$
In this case
$$
v(y)=\frac{8}{3}gb(2a-3b)e^{-2by}+16g^2b^2e^{-4by}+\frac{4a^2}{9}.
$$
After the DT
$$
v\to v^{(1)}=v-4\frac{d^2}{dy^2}\log A,
$$
we get
$$
v^{(1)}(y)=\frac{8}{3}gb(2a+3b)e^{-2by}+16g^2b^2e^{-4by}+\frac{4a^2}{9}.
$$
Thus $v^{(1)}$ can be obtained from $v$ by substitution $a\to -a$
and $g\to -g$. It means that $v(y)$ is an SP-potential.
\newline
{\em C. Odd superpotential.} In this case we have
$$
\log A(y)=-ay-by^2.
$$
A calculation yields
$$
v(y)=4\left(2by+a\right)^2-4b.
$$
It is nothing else, but the harmonic oscillator and, hence, an
SP-potential: $ v^{(1)}=v+{\rm const}$. Let's consider this
example as the fuse in further constructing of the new exact
solutions via DT.

\subsection{\label{sec:level4}The generalization of the odd superpotential: DT on the harmonic oscillator
background}

As we have seen, the odd superpotential $W(\phi(y))=2\kappa
b^2\phi(y)$, with $\kappa=+1$ ($y>0$) or $\kappa$=-1 ($y<0$)
results in
\begin{equation}
\begin{array}{cc}
\phi(y)=\phi_0+\kappa b^2y,\\
\\
\log A(y)=\log A_0+\frac{\phi^2_0-\phi^2(y)}{3}, \label{op}
\end{array}
\end{equation}
where $A_0=A(0)$, $\phi_0=\phi(0)$. Using (\ref{op}) we get the
Schr\"odinger equation of the harmonic oscillator
$$
v(y;\kappa)=\frac{16}{9}b^4\left(\kappa b^2y+\phi_0\right)^2.
$$
\newline
Its ground state has the wave function $\psi_1(y;\kappa)={\rm
e}^{-2(\kappa b^2y+\phi_0)^2/3}$, and energy $\lambda_1=4b^4/3$.
Since $\psi_1(0;+1)=\psi_1(0;-1)$ and
$$
{\dot\psi}_1(0;+1)-{\dot\psi}_1(0;-1)=-\beta{\rm
e}^{-2\phi^2_0/3},
$$
the tension will be $\beta=8b^2\phi_0/3$. All other eigenfunctions
and eigenvalues can be obtained via the formulas
\begin{equation}
\begin{array}{cc}
 \psi_n(y;\kappa)=\left(\partial_y-4 b^2(b^2
y+\kappa\phi_0)/3\right)^{n-1}\psi_1(y;\kappa),\\
\\
\lambda_n=4b^2(2n-1)/3. \label{psin}
\end{array}
\end{equation}
In the article \cite{Adler} Adler has suggested the more general
way of deletion of a groups of an excited states. It ensures the
general state $\lambda_{_{vac}}$ of a transformed Hamiltonian to
stay the same, though being described by a totally new wave
function. Using this theorem and (\ref{crum}) one can delete the
even number of adjacent levels (if the number will be odd, for
example N=1 like before, then the new potential appears to be
singular). Upon deletion of levels $\lambda_3$ and $\lambda_4$ one
gets Higgs-like potential $v^{(2)}(y;\kappa)$, with the ground
state $\psi^{(2)}_1(y;\kappa)$ and the tension $\beta^{(2)}$:

\begin{widetext}
\begin{equation}
\begin{array}{cc}
\displaystyle{v^{(2)}=\frac{16b^4}{9}\frac{4096x^{10}+12288
x^8+21888 x^6+10368 x^4-22599 x^2+2187}{(64 x^4+27)^2}, \qquad
\psi^{(2)}_1=\frac{8x^2+3}{64x^4+27}{\rm e}^{-2x^2/3}}\\
\\
\displaystyle{\beta^{(2)}=\frac{8b^2\phi_0(512\phi^6_0+960\phi^4_0+792\phi^2_0-243)}{3(64\phi^4_0+27)(8\phi^2_0+3)}},
\label{v2}
\end{array}
\end{equation}
\end{widetext}
where $x=\kappa b^2y+\phi_0$. The potential $v^{(2)}$ is
represented on the Figure 1.

\begin{figure}
\centering\leavevmode\epsfysize=4.5cm \epsfbox{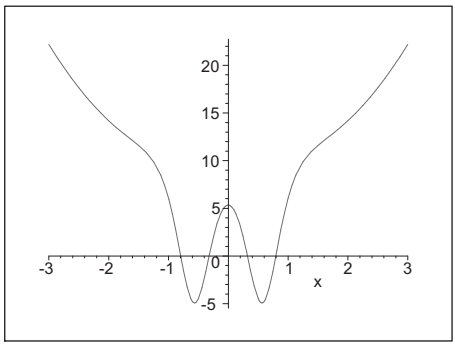}
\caption{\label{fig:epsart} The harmonic oscillator without levels
$\lambda_2$ and $\lambda_3$.}
\end{figure}

After the deletion of levels $\lambda_3$, $\lambda_4$, $\lambda_6$
and $\lambda_7$ we receive one very interesting potential which is
shown on Figure 2. Finally, the Figure 3 represents the potential
which is obtained through the elimination of the levels
$\lambda_4-\lambda_7$.
\begin{figure}
\centering\leavevmode\epsfysize=4.5cm \epsfbox{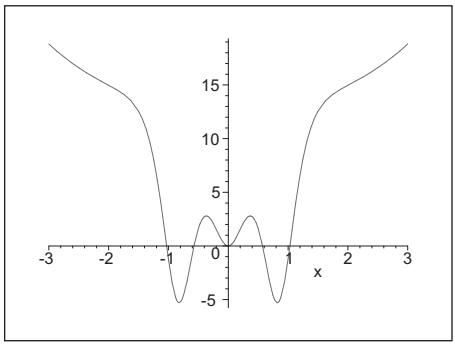}
\caption{\label{fig:epsart}The harmonic oscillator without levels
$\lambda_3$, $\lambda_4$, $\lambda_6$, $\lambda_7$.}
\end{figure}
The expressions for the potentials $v^{(n)}(y;\kappa)$,
eigenfunctions (including the ground state) and tensions
$\beta^{(n)}$ can be easily obtained via the (\ref{crum}) and
(\ref{psin}). We omit them only for their extreme bulkiness.

The reason why these potentials have such Higgs-like form is
evident. In fact, the spectrum of potential with two symmetric
minimums necessarily contains the lacuna between two down levels
and all other spectrum. Deleting the levels $\lambda_3$ and
$\lambda_4$ we construct the spectrum with such lacuna so it is
only natural that this procedure results in nothing else but
Higgs-like potential. In a similar manner, elimination of the
levels $\lambda_4-\lambda_7$ leads to the spectrum with lacuna
between the three bottom levels and the other spectrum. As a
result one get the potential with three minimums (Fig. 3). Note
that all these potentials have the following asymptotic:
$v^{(n)}(y;\kappa)\to 16b^8y^2/9$, when $y\to\pm\infty$.

Now lets return to potential (\ref{v2}). Using the exact forms of
the $v^{(2)}$ and $\psi_1^{(2)}$ one can obtain the potential
$V(\phi)$ in parametric form (i.e. $V=V(y)$ and $\phi=\phi(y)$).
The plot of the function $d\phi(x)/dx$ for $y>0$ is represented on
the Fig. 4. This expression is real for $\mid x\mid\ge x_*\sim
0.25735$ so if we identify the $\phi_0=x_*$ then the field
$\phi(y)$ will be well determined all around the bulk.


\begin{figure}
\centering\leavevmode\epsfysize=4.5cm \epsfbox{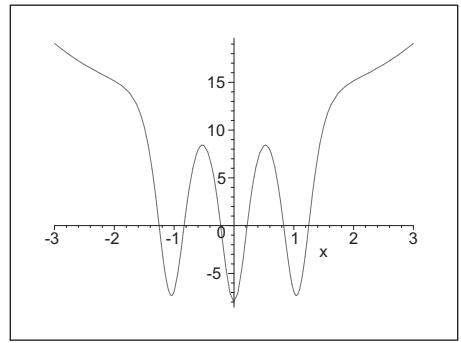}
\caption{\label{fig:epsart}The harmonic oscillator without levels
$\lambda_4-\lambda_7$.}
\end{figure}
\begin{figure}
\centering\leavevmode\epsfysize=4.5cm \epsfbox{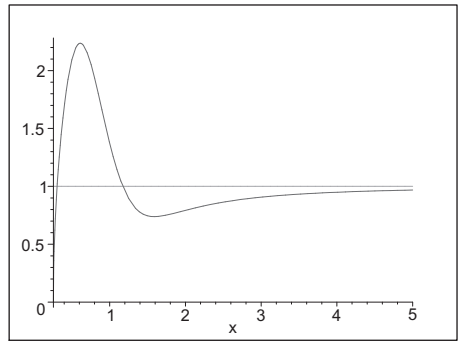}
\caption{\label{fig:epsart}The function $\phi'(x)$.}
\end{figure}

\subsection{\label{sec:level5}Dressing chains and the fourth Painlev\'e equation as
the generalization of the odd superpotential}

Another interesting generalization of the odd superpotential (i.e.
the harmonic oscillator) can be obtained with the aid of the
dressing chains of the DT, which chains were previously introduced
in \cite{cepi}. Let's suppose, that we have defined the set of
functions $f_k=\partial_y\log\psi_k^{(k-1)}$, where $\psi_k^{(n)}$
are $n$-times dressed solutions of the (\ref{schrodinger}). Then
these functions will also be solutions of the {\em dressing
chain}:
\begin{equation}
{\dot f}_{k}+{\dot f}_{k+1}=f_k^2-f_{k+1}^2+\alpha_k.
\label{chain}
\end{equation}
where $\alpha_k$ are constants, and they can be expressed via
spectral parameters $\lambda_k$. Following \cite{cepi} we'll
consider the periodic version $f_k=f_{k+N}$,
$\alpha_k=\alpha_{k+N}$ with positive integer $N$ which will be
further referred as period. The whole theories of dressing chains
are totally different for the odd ($N=2n+1$) and even ($N=2n$)
periods. The same is valid for the cases with $\alpha=0$ and
$\alpha\ne 0$, where
$$
\alpha=\sum_{k=1}^N\alpha_k.
$$
If $\alpha\ne 0$ and $N=1$ we get the harmonic oscillator. In the
case $N=3$ ($\alpha\ne 0$) we get the fourth Painlev\'e equation
($P_{_{IV}}$) for the function $P(y)=-f_1-y$:
\begin{equation}
{\ddot P}=\frac{{\dot
P}^2}{2P}+\frac{3}{2}P^3+4yP^2+2\left(y^2-a\right)P+\frac{b}{P},
\label{Painleve}
\end{equation}
where $\alpha=2$, $a=-\alpha_1-\alpha_2/2+1$, $b=-\alpha_2^2/2$.
Thus the potential $v(y)={\dot f}_1+f_1^2$ can be expressed in
term of the Painlev\'e-$IV$ transcendents and has the following
behavior
\begin{equation}
 v(y-y_0)\sim
\frac{\alpha^2y^2}{36}+My\cos\left(\frac{\alpha
y^2}{\sqrt{3}}+\delta_0\right). \label{vv}
\end{equation}
Using (\ref{vv}) one can conclude that this potential is indeed
the generalization of the harmonic oscillator. In particular, this
potential is growing quadratically.

We should also mention, that the spectrum of the $v(y)$ can be
found explicitly \cite{cepi}. Moreover, it can be shown that the
ground state is at $\lambda_1=0$.

\section{The models with cosmological expansion}

If we are going to study the models with cosmological expansion,
we should expose the DT method to some kind of generalization.
This requirement inevitably follows from the fact, that the
substitution $A^2=\psi$ now results in nonlinear equation rather
then in linear Schr\"odinger equation. However, seemingly being as
difficult as it seems, this equation still can be reformulated in
the way of linear spectral problem which does admit the Darboux
transformation. In the next subsection we'll consider the
spacetimes which are compact in $y$. We'll then introduce the
simple orbifold $S_1/{\Bbb Z}_2$ and will show that, upon usage of
DT, one can receive the solutions which result in exponentially
small effective 4-D cosmological constant. In the last subsection
we'll briefly consider the new method which is valid for models
both with and without cosmological expansion.

\subsection{\label{sec:level2} Noncompact spacetimes}

If in (\ref{interval}) $H\ne 0$ then instead of (\ref{dvaurav}) we
have a much more complex system:
\begin{equation}
\begin{array}{c}
\displaystyle{ {\dot
u}=-\frac{2{\dot\phi}^2}{3}-\frac{2H^2}{A}-\frac{2}{3}\sum_{b=1}^N\sigma_b\delta(y-y_b)},\\
\displaystyle{
u^2=\frac{{\dot\phi}^2}{3}-\frac{2}{3}V(\phi)+\frac{4H^2}{A},}
\label{Hne0}
\end{array}
\end{equation}
written for the set of $N$ branes which are located at $y=y_b$.

The function $A^2$ solves the following nonlinear equation
\begin{equation}
\left(-\frac{d^2}{dy^2}-\frac{8}{3}V(\phi(y))+\frac{12
H^2}{A}-\frac{4}{3}\sum_{b=1}^N\sigma_b\delta(y-y_b)\right)A^2=0.
\label{A2}
\end{equation}
Therefore, we can formally introduce the spectral problem (the
Schr\"odinger equation):
\begin{equation}
{\ddot\psi(y;\lambda)}=\left(U(y)-\sum_{b=1}^N\beta_b\delta(y-y_b)-\lambda\right)\psi(y;\lambda),
\label{nschrodinger}
\end{equation}
 where
\begin{equation}
\displaystyle{
U(y)=-\frac{8}{3}V(\phi(y);\lambda)+\frac{12H^2}{\sqrt{\psi(y;\lambda)}}.}
\label{U}
\end{equation}
We stress here, that $U(y)$ is not a function of the spectral
parameter $\lambda$ whereas $V$ and $\psi$ are. With this induced,
we are free to use the spectral theory for the equation
(\ref{nschrodinger}). For any given $U(y)$ one can solve
(\ref{nschrodinger}) to get $\psi$ and to find $V$ from (\ref{U}).
Of course, at the end of our solution we should, in order to find
$A=\sqrt{\psi}$, receive $\psi$ of positive value, but this
problem is also avoidable, as has been demonstrated in sections 3
and 4. Indeed, one can construct positive solution via the DT
starting out from any simple solvable model - taking the RS model
for example. There is, however, one interesting difference with
the case of stationary brane: the ''kinetic term'' appears to have
the form
\begin{equation}
\frac{{\dot\phi}^2}{3}=\frac{1}{4}\left(\left(\frac{\dot\psi}{\psi}\right)^2-U+\lambda\right)-\frac{H^2}{\sqrt{\psi}},
\label{kinet}
\end{equation}
therefore we can't just use the ground state solution like we did
in the case of stationary brane, or at $y\to\infty$ wave function
$\psi\to 0$ and we will get ${\dot\phi}^2<0$. Obviously, this
problem requires a more through examination. So, let's again start
out from an ``RS potential'' $U=\mu^2$, and consider the case
$\lambda=0$. The solution will have the form (\ref{opor}). Using
it as the prop function in (\ref{dt}), one gets the same metric
$A^{(1)}_{\pm}$ from section 3, only with another potential. The
main problem is now focused in the kinetic energy of the scalar
field, having the form
$$
\displaystyle{
({\dot\phi^{(1)}_{\pm}})^2=\frac{3\mu^2a_{_{\pm}}b_{_{\pm}}}{\left(a_{_{\pm}}e^{\mu
y}+b_{_{\pm}}e^{-\mu y}\right)^2}-3H^2\sqrt{a_{_{\pm}}e^{\mu
y}+b_{_{\pm}}e^{-\mu y}}.}
$$
It is clear that for large enough $y$, one gets $\dot\phi^2<0$. On
the other hand, if (see (\ref{represent}))
$$
\left(\frac{H}{\mu}\right)^2<\frac{\sin^22\alpha_{\pm}}{4r},
$$
then on the brane and in its vicinity, there are no such troubles.
Therefore the simplest way of avoiding the problem is just dealing
with the finite volume, which is less then the minimum value of
$y$, resulting in the negativity of the kinetic term.

If we nevertheless wish to deal with infinite volume, we had to be
sure to use only those positive solutions
$\psi(y)=\{\psi_{+}(y),\,\,{\rm at}\,\,y>0; \psi_{-}(y),\,\,{\rm
at}\,\,y<0\}$ which are growing as $\mid y\mid \to \infty$. The
construction of such solutions can be done in different manners.
For example, one can take the solution of the Schr\"odinger
equation with $\lambda=\lambda_{-1}<\lambda_{_{vac}}$. The more
interesting opportunity lies in applying of so called B-potential
(Bargmann potentials) $U(y)$ whose solutions of the Schr\"odinger
equation all have the form: $\psi=F(y)\exp(\alpha y)$ where $F(y)$
is some polynomial \cite{Bargmann}. To show them in work lets
rewrite the equation (\ref{kinet}) in the form
\begin{equation}
\frac{1}{3}{\dot\phi}^2_{_{\pm}}=-\frac{1}{2}{\dot
u}_{_{\pm}}-\frac{H^2}{\sqrt{\psi_{_{\pm}}}}, \label{kinet1}
\end{equation}
where $u_{_{\pm}}={\dot\psi}_{_{\pm}}/(2\psi_{_{\pm}})$. The
necessary criterion here is that ${\dot u}_{_{\pm}}<0$ for any
given $y$. Let's choose the $\psi_{_{\pm}}$ in typical ''Bargmann
form'':
\begin{equation}
\psi_{_{\pm}}=\left(Ay^2\pm B y+C\right){\rm e}^{\pm\alpha y},
\label{Barg}
\end{equation}
where $\alpha$, $A$, $B$ and $C$ are positive constants,
associated by the condition
\begin{equation}
2AC<B^2<4AC. \label{B2}
\end{equation}
The jump condition (\ref{jump}), imposed to single brane at $y=0$
results in $\beta=-2(B+\alpha C)/C$ . The solution of the equation
(\ref{Hne0}) appears to be
\begin{equation}
A_{_{\pm}}={\rm e}^{\pm \alpha y /2} \sqrt{Ay^2\pm B y+C}.
\label{Apm}
\end{equation}
where $A(y)=A_{_+}$ at $y>0$ and $A(y)=A_{_-}$ at $y<0$. The
condition (\ref{B2}) guarantees that this solution will be
nonsingular on brane as well as in bulk (since $A(y)\ne 0$ for any
values of $y$) and also that ${\dot u}_{_{+}}<0$ for $y>0$ and
${\dot u}_{_{-}}<0$ for $y<0$. Finally, choosing
$$
H^2<\frac{B^2-2AC}{4C\sqrt{C}},
$$
and using (\ref{Apm}) and (\ref{kinet1}) one get
$$
\frac{{\dot\phi}^2_{_{\pm}}}{3}=\frac{2A^2y^2\pm
2ABy+B^2-2AC}{4(Ay^2\pm By+C)^2}-\frac{H^2{\rm e}^{\mp\alpha y /
2}}{\sqrt{Ay^2\pm By+C}},
$$
where $\phi(y)=\phi_{_+}$ at $y>0$ and $\phi(y)=\phi_{_-}$ at
$y<0$. Now we can see, that it is always possible to find large
enough values of $\alpha$ that will greatly decrease the second
member of equation with regard to the first, thus leaving the
whole equation positive for both $y>0$ and $y<0$.
\subsection{\label{sec:level3}Orbifold model}

In previous sections, our efforts were wholly concentrated on
spacetimes that are noncompact in $y$. Now we'll consider the
compactified case. As we shall see, the DT in this case allows to
obtain the solutions, resulting in exponentially small 4-D
effective cosmological constant.

Let's choose the potential in (\ref{nschrodinger}) as
$U=\mu^2\sin^2\theta$. We'll start with the two solution of this
equation:
$$
 \psi_{_1}=\cosh(\mu\sin\theta y),\qquad
\psi_{_{2\pm}}=\sinh(\mu y\pm b),
$$
\newline
where $b=$const. The function $\psi_{_1}$ is the solution of the
(\ref{nschrodinger}) with $\lambda_1=0$ whereas $\psi_{_{2\pm}}$
with $\lambda_2=-\mu^2\cos^2\theta$. Using $\psi_{_1}$ as the prop
function we can dress $\psi_{_{2\pm}}$ with the help of the DT
(\ref{dt}). The result will be
\begin{equation}
\psi^{(1)}_{_{2\pm}}=\cosh(\mu y\pm
b)-\sin\theta\tanh(\mu\sin\theta y)\sinh(\mu y\pm b). \label{oppa}
\end{equation}
where we should choose the sign ''$+$'' for $y>0$ and sign ''$-$''
for $y<0$. One can show that $\psi^{(1)}_{_{2\pm}}>0$ for all $y$.
In this case the dressed potential $U^{(1)}$ will have the single
bound state with $\lambda=\lambda_1$ and eigenfunction
$1/\psi_{_1}$.

Now we can compactify the $y$ direction as an ${\Bbb Z}_2$
orbifold, with two branes sitting at each of the two fixed points
($y = 0$ and $y=L$). The new aspect here is that $S_1/{\Bbb Z}_2$
orbifold model permits only one bulk space. Due to the symmetry of
the model, we only have to consider the jump conditions at $y = 0$
and $y = L$. The direct calculations result in tension
$\sigma_0=-3\mu \tanh b\cos^2\theta/2$
for the brane which is located at $y=0$ (we'll choose it as the
visible brane) and

\begin{widetext}
$$
\sigma_{_L}=-\frac{3\mu}{2}\frac{\sinh(\mu
L+b)(\cosh^2(\mu\sin\theta
L)-\sin^2\theta)-\sin\theta\sinh(\mu\sin\theta
L)\cosh(\mu\sin\theta L)\cosh(\mu L+b)}{\cosh(\mu\sin\theta
L)\left(\sin\theta\sinh(\mu\sin\theta L)\sinh(\mu
L+b)-\cosh(\mu\sin\theta L)\cosh(\mu L+b)\right)}, \label{sigmal}
$$
\end{widetext}
as a tensions on $y=L$ branes. These expressions can be greatly
simplified by the choice $b=-\mu L$:
$$
\begin{array}{cc}
\displaystyle{
\sigma_0=\frac{3\mu}{2}\tanh(\mu L)\cos^2\theta,}\\
\\
\displaystyle{
 \sigma_{_L}=-\frac{3\mu}{2}\tanh(\mu\sin\theta
L)\sin\theta.}
\end{array}
$$
Since the expansion rate on the visible brane is $H(0) =
H/\sqrt{A(0)}=H/\left(\psi^{(1)}_{_{2\pm}}(0)\right)^{1/4}$ then
$$
H(0)=\frac{H}{{}^4\sqrt{\cosh(\mu L)}}.
$$
Therefore, for large values of $\mu L$ the effective 4-D
cosmological constant $\Lambda_{eff}$ on the visible brane is
\begin{equation}
\Lambda_{eff}\sim H^2(0)=\frac{H^2}{\sqrt{\cosh(\mu L)}}\sim
H^2{\rm e}^{-\mu L/2}. \label{Lambda0}
\end{equation}
This means, that 4-dimensional cosmological constant, as seen by
observers from the visible brane, becomes exponentially small
while the $L$ grows large, therefore, automatically solving the
cosmological constant problem right in the framework of the the
brane worlds models (this idea was first suggested in article of
S.-H.H. Tye and I. Wasserman \cite{TW}).

All we need now is to make sure, that ${\dot\phi}^2_{_{+}}>0$ at
$0<y<L$ (since we have the single bulk between two branes then it
is enough to consider the case $0<y<L$ solely). Let's choose
$\theta\sim\pi/2$. In this case $\lambda_2\sim\lambda_1=0$. We
stress, however, that $\theta\ne \pi/2$ exactly so
$\lambda_2=-\mu^2\alpha^2$ where $\alpha=\pi/2-\theta\ll 1$. To
obtain ${\dot\phi}^2_{_{+}}(0)>0$ it is necessary that $\cosh(\mu
L)>16 H^4/\mu^4$ whereas ${\dot\phi}^2_{_{+}}(L)>0$ if $\cosh(\mu
L)<\mu/(2H)$. The combination of these conditions is:
$$
\left(\frac{2H}{\mu}\right)^4<\cosh(\mu L)<\frac{\mu}{2H},
$$
which is possible if $\mu>2H$. In fact, it is better to imply an
enhanced condition $\mu\gg 2H$, because large values of $\mu$
allows to avoid the conclusion that $\sigma_0=3\mu\alpha^2/2\sim
0$. Since $\mu$ is not bounded from above, we draw the wanted
conclusion, that our model indeed permits positive
${\dot\phi}^2_{_{+}}$ at $0<y<L$.
\newline
\newline
{\bf Nota Bene 3}. Of course, the described theory could not
generally protect some finite number of regions (at $0<y<L$) from
special situations when ${\dot\phi}^2_{_{+}}(y)<0$ (if the value
of $\mu$ is not sufficiently large, for example). In this case,
however, one can use the method which was supposed in
\cite{Yurov}, viz. these regions can be cut out, with their
boundaries being sewn together in such a way that neither the warp
factor (along with its first two derivatives) nor Ricci scalar
will experience a jump. This unexpected fortune raises from the
fact that matching is done at an inflection point of $\log A$
(more correctly, this is approximately true if $\psi$ is
sufficiently large to neglect the term $H/\sqrt{\psi}$ in either
one of equations (\ref{kinet}) and (\ref{kinet1})).


\subsection{\label{sec:level4}Green function method}

This is another way to construct solution for the models with
cosmological expansion. First of all, we can directly transform
the system (\ref{Hne0}) into the linear equation rather than use
substitution $\psi =A^2$. Indeed, introducing the potential
$$
\textrm{U}\equiv
-\frac{2}{3}\left(\frac{{\dot\phi}^2}{2}+V(\phi)\right),
$$
we get (for the case of single brane located at $t=0$)
\begin{equation}
-{\ddot
A}+\left(\textrm{U}-\frac{2}{3}\sigma_0\delta(y)\right)A=-2H^2.
\label{sr-2}
\end{equation}
Solving linear nonhomogeneous equation (\ref{sr-2}) for given
$\textrm{U}$ we can find $\phi(y)$ and $V(\phi)$ via the system
$$
\begin{array}{cc}
\displaystyle{{\dot\phi}^2=-\frac{3}{2}\left(\textrm{U}-\left(\frac{\dot
A}{A}\right)^2\right)-\frac{6H^2}{A}},\\
\displaystyle{V=-\frac{3}{4}\left(\textrm{U}+\left(\frac{\dot
A}{A}\right)^2\right)+\frac{3H^2}{A}}.
\end{array}
$$
Now, instead of the (\ref{sr-2}) let's consider the nonhomogeneous
spectral problem
$$
-{\ddot\psi}(y;\lambda)+\left(\textrm{U}(y)-\lambda-\frac{2}{3}\sigma_0\delta(y)\right)\psi(y;\lambda)=-2H^2.
$$
which can be written as
\begin{equation}
{\bf L}(y;\lambda)\psi(y;\lambda)=-2H^2, \label{NS}
\end{equation}
with ${\bf
L}(y;\lambda)=d^2/dy^2+U(y)-\lambda-2\sigma_0\delta(y)/3$.

To solve (\ref{NS}) we introduce the homogeneous linear equation
\begin{equation}
 {\bf L}(y;\lambda)\psi_0(y;\lambda)=0,
 \label{Odn}
\end{equation}
and the Green function $G(y,z;\lambda)$:
$$
 {\bf L}(y;\lambda)G(y,z;\lambda)=\delta(y-z).
$$
By solving these equation for given $U(y)$ and $\lambda$ we can
construct the solution of (\ref{NS}) as
$$
\psi(y;\lambda)=\psi_0(y;\lambda)-2H^2\int dz\, G(y,z;\lambda).
$$
If $\psi(y;\lambda)$ will be positive then we will be able to
define $A=\sqrt{\psi}$ just like before. But first we have to
determine those conditions which make such positive solutions
possible. This is still the open question which will be considered
in the next publications. We should only note, that Green function
can be constructed via solutions of the (\ref{Odn}). If initial
potential $\textrm{U}(y)$ is exactly solvable one, then the same
is true for N-times Darboux dressed potential $\textrm{U}^{(N)}$.
Thus, if we know the Green function $G(y,z;\lambda)$ of the
initial equation then we know the Green function
$G^{(N)}(y,z;\lambda)$ for equation with the dressed potential. In
other words, the DT is appliable here, at least formally.

\section{Conclusion}
This approach can be easily generalized to the case of $K$-branes.
For this we should replace the term $\beta\delta(y)$ in
(\ref{schrodinger}) by
$$
\sum_{b}\beta_b\delta(y-y_b),
$$
and also replace the two jump conditions (\ref{jump}) by $2K$ jump
conditions at $y=y_b$. This is nothing more than technical matter,
that's why we omit the details.

Another one kettle of fish is that even in case of general
position, where we cannot reconstruct the form of the potential
$V(\phi)$ as a function of $\phi$, all these solutions will have
the RS asymptotics (of course, if constructed precisely according
to section 3). The DT gives a link between the solvable problems,
and probably most (if not all) of the exactly solvable potentials,
from the potential of the harmonic oscillator to the finite-gap
potentials \cite{cepi}, can all be obtained via these
transformations. The physical sense of these potentials in the
brane world is not clear. But our main purpose in this work was to
merely demonstrate and advertise the DT as being the very powerful
tool for manufacturing the exactly solvable potentials in the 5-D
gravity with the bulk scalar field. It has been demonstrated that
this method can be useful both in models with and without the
cosmological expansion, although in the first case the situation
is more complex and deserves the further studying. Notably, there
is one imperfection of this method: the DT would work for the 5-D
gravity only. If $D>5$ then the warp factor is a multi-variable
function. Unfortunately, there is no cogent and effective general
DT theory in several dimensions \cite{D-Moutard}. But,
nevertheless, for $D=5$  the DT is by far seems to be the best way
to construct the nonsingular exact solutions.
\begin{acknowledgments}
The authors are grateful to D. Vassilevich for useful comments,
the Helmholtz program for financial support and  M. Bordag for his
kind hospitality at the University of Leipzig. Research has been
partially supported by "Integration"  Grant N$\Phi$ 0032/1242.
\end{acknowledgments}
$$
{}
$$

\bibliography{apssamp}

\centerline{\bf References} \noindent
\begin{enumerate}
\bibitem{Flanagan} E.E. Flanagan, S.-H.H. Tye and I. Wasserman, \rm\, Phys. Lett. {\bf B 522}, (2001), 155-165, [hep-th/0110070 v2].
\bibitem{uskor}
If the observed cosmological accelerated expansion  is generated
by the nonzero cosmological constant, the 4-D metric in
(\ref{interval}) can, in fact, describe our universe - of course,
in case of future distant enough. See S. Perlmutter et al.,
Astrophys. J. 517, 565 (1999) [astro-ph/9812133]; A. G. Riess et
al., Astron. J. 116, 1009 (1998) [astro-ph/9805201]; P. M.
Garnavich et al., Astrophys. J. 509, 74 (1998) [astro-ph/9806396].
\bibitem{RS} L. Randall and R. Sundrum, \rm\, Phys. Rev. Lett. {\bf 83}, 3370 (1999), [hep-th/9906064].
\bibitem{TW} S.-H.H. Tye and I. Wasserman,\rm\, Phys. Rev. Lett. {\bf 86} 1682 (2001) [hep-th/0006068].
\bibitem{DFGK} O. DeWolf, D.Z. Freedman, S.S. Gubser and A. Karch,\rm\, Phys. Rev. {\bf D62}, 046008 (2000), hep-th/9909134.
\bibitem{kaku} Z. Kakushadze, \rm Nucl. Phys. {\bf B 589}, 75 (2000) [hep-th/0005217].
\bibitem{Salle}  V.B. Matveev and M.A. Salle.\rm\,  Darboux Transformation and
    Solitons. Berlin--Heidelberg: Springer Verlag, 1991.
\bibitem{ABI} A.A. Andrianov, N.V. Borisov and M.V. Ioffe,\rm\,Phys.Lett.{\bf A(105)}
19 (1984); A.A. Andrianov, N.V. Borisov M.I. Eides and M.V.
Ioffe,\rm\,Phys.Lett.{\bf A(109)} 143 (1984).
\bibitem{Israel} W. Israel, Nuovo Cimento {\bf B 44} (S10) 1 (1966); V.A. Berezin, V.A. Kuzmin and I.I. Tkachev, \rm
Phys. Rev. {\bf D36}, 2919 (1987).
\bibitem{1} R.R. Caldwell, M. Kamionkowski and N.N. Weinberg, [astro-ph/0302506].
\bibitem{2} R.R. Caldwell, Phys. Lett. B 545, 23-29 (2002) [astro-ph/9908168].
\bibitem{3} S.M. Carroll, M. Hoffman and M. Trodden,  [astro-ph/0301273].
\bibitem{4} H.-P. Nilles, Phys. Rep. 110, 1-162 (1984).
\bibitem{5} M.D. Pollock, Phys. Lett. B 215, 635-641 (1988).
\bibitem{7} V. Sahni and Y. Shtanov, [astro-ph/0202346].
\bibitem{Darboux} J.G. Darboux \rm\, Compt.Rend., 94 p.1343 (1882).
\bibitem{bereza} B.P. Berezovoy, A.I. Pashnev, Theor. and Math. Phys., v.70, N1, 146  (1987).
\bibitem{Rubakov} V.A. Rubakov, Uspehi Fiz. Nauk {\bf 171}, 913 (2001).
\bibitem{Crum} M.M. Crum \rm\, Quart. J. Math. Ser. Oxford 2, 6 p. 121
(1955).
\bibitem{thereby} Thereby, defining the integration's constants $C_1=1$ and $C_2=0$, which is
distinct for a ground state.
\bibitem{Infeld} L. Infeld and T.E. Hull, \rm \, Rev.Mod.Phys.  V. 23 p.
21 (1951).
\bibitem{Adler}  V.E. Adler, Theor. and Math. Phys., v.101, N3, 323
(1994).
\bibitem{Bargmann} As a matter of fact, in the case of  Bargmann potentials
the solutions of the Schr\"odinger equation have the form
$P(k)\exp(iky)$, where $P(k)$ is some polynomial. We use some
generalization of B-potentials. See V. Bargmann, Rev. Mod. Phys.
21, 488-493 (1949).
\bibitem{Yurov} A.V. Yurov, S.D. Vereshchagin,
Theoretical and Mathematical Physics, 139 (3): 787-800 (2004);
A.V. Yurov, [astro-ph/0305019].
\bibitem{cepi} A.P. Veselov and A.B. Shabat,\rm\, Funkts. Anal. Prilozhen.,
Vpl. 27, No. 2, p. 1 (1993).
\bibitem{D-Moutard} A. Gorizales-Lopes and N. Kamran, J.Geom.Phys.{\bf 26}202-226 (1998) [hep-th/9612100].
\smallskip
\end{enumerate}

\end{document}